\documentclass[aps,twocolumn,superscriptaddress,preprintnumbers,showpacs]{revtex4}
\usepackage[colorlinks,linkcolor=blue,urlcolor=blue,citecolor=blue,bookmarks,bookmarksnumbered]{hyperref}
\usepackage{bm}
\usepackage{dcolumn}
\usepackage{graphicx}
\usepackage{booktabs}
\usepackage{multirow}
\usepackage{setspace}
\usepackage{natbib}
\usepackage{cases}

\newcommand{\e}{{\mathrm{e}}}
\renewcommand{\i}{{\mathrm{i}}}
\renewcommand{\deg}{^\circ}

\begin{document}

\newcommand*{\PKU}{School of Physics and State Key Laboratory of Nuclear Physics and
Technology, Peking University, Beijing 100871,
China}\affiliation{\PKU}
\newcommand*{\CIC}{Collaborative Innovation Center of Quantum Matter, Beijing, China}\affiliation{\CIC}
\newcommand*{\CHEP}{Center for High Energy Physics, Peking University, Beijing 100871, China}\affiliation{\CHEP}

\title{Neutrino properties from ultra-high energy cosmic neutrinos}

\author{Yanqi Huang}\affiliation{\PKU}
\author{Bo-Qiang Ma}
\email{mabq@pku.edu.cn}\affiliation{\PKU}\affiliation{\CIC}\affiliation{\CHEP}

\begin{abstract}
 Neutrino properties can be constrained by the detection of ultra-high energy cosmic neutrinos (UHECNs). By using the updated global fitting results of neutrino mixing parameters, we present predictions on the neutrino flavor ratios at the Earth from three possibly astrophysical sources. Comparing with the latest IceCube data, we find that the normal hierarchy (NH) and inverted hierarchy (IH) cases from the initial ratios $\phi_{\nu_e}^0:\phi_{\nu_{\mu}}^0:\phi_{\nu_{\tau}}^0=$1:2:0 and 0:1:0 are compatible with the data in the standard neutrino oscillation scenario. We also examine the neutrino flavor ratios in a neutrino decay scenario beyond the standard model, and introduce the special case that two mass eigenstates
 of neutrinos, i.e., $\nu_1$ and $\nu_2$, are degenerated. We find that the IH case and the degenerate NH case from the 1:2:0 and 0:1:0 sources are still permissible with the IceCube data within the $3\sigma$ error range. The general constraints only rely on the neutrino mixing and oscillation framework are also discussed.

\end{abstract}


\maketitle


The properties of neutrinos are under active investigations from various experiments.
The neutrino oscillation has been well established by the experiments with solar, atmospheric, reactor and accelerator neutrinos in recent decades~\cite{e1,e2,e3,e4,e5,e6,e7,e8,e9}.
There have been a number of new measurements on the neutrino mass splitting recently~\cite{MINOS,Daya Bay,T2K new,zhang}. However, the hierarchy of mass eigenstates is still undecided, and it is also not clarified yet whether neutrinos are Dirac or Majorana fermions.

The detection of ultra-high energy cosmic neutrinos
(UHECNs) can provide us information concerning neutrino properties~\cite{astrophysical neutrino decay, astrophysical neutrino}. There are many running or on-going experiments such as ANITA~\cite{ANITA}, ARA~\cite{ARA}, and IceCube~\cite{IceCube}. The IceCube collaboration just reported the latest results of ultra-high energy cosmic neutrino flavor ratios~\cite{IceCube new1, IceCube new2}. The UHECNs with energy above $10^7$~eV are expected to arise from a series of processes in the ultra-high energy cosmic rays (UHECRs).  One of the components of the UHECRs is the ultra-high energy proton flux, which is thought to origin from the extra-galactic processes such as gamma-ray bursts (GRBs) or active galactic nuclei (AGNs)~\cite{UHECRs}. These accelerated protons may scatter with the intergalactic medium or the cosmic microwave background~(CMB) photons and generate the pions:
 \begin{eqnarray}
p+p&\rightarrow &\pi^+\pi^-\pi^0,\\ p+\gamma&\rightarrow&\Delta^+\rightarrow n+\pi^+.\label{GZK}
\end{eqnarray}
  The proton-photon scattering in Eq.~(\ref{GZK}) is known as the Greisen-Zatasepin-Kuzmin~(GZK) process~\cite{GZK1,GZK2}. The charged pions and their muon daughters generate the UHECNs by the decay chains
  \begin{eqnarray}
  \pi^+&\rightarrow&\mu^++\nu_{\mu}\rightarrow e^++\nu_{\mu}+\bar{\nu}_{\mu}+\nu_e,\label{pion decay 1}\\ \pi^-&\rightarrow&\mu^-+\bar{\nu}_{\mu}\rightarrow e^-+\nu_{\mu}+\bar{\nu}_{\mu}+\bar{\nu}_e,\label{pion decay 2}
  \end{eqnarray}
 which result in the most conventional initial ratios in the pion-decay case~\cite{pion decay 1, pion decay 2}
  \begin{equation}
  \phi_{\nu_e}^0:\phi_{\nu_{\mu}}^0:\phi_{\nu_{\tau}}^0=1:2:0.\label{initial ratio}
  \end{equation}
  Besides the pion decays, some other generative mechanisms of UHECNs were proposed. If the muons lost energy while they are going through strong magnetic fields or matters~\cite{pure muon 1, pure muon 2}, the energy of $\nu_e$ and $\bar{\nu}_e$  emitted in the secondary decays in Eqs.~(\ref{pion decay 1}) and (\ref{pion decay 2}) are lower than the ultra-high energy range we concern. Hence, at
high energy the $\nu_{e}$ flux is decreased and the ratios approximately change into the muon-damped case:
  \begin{equation}
  \phi_{\nu_e}^0:\phi_{\nu_{\mu}}^0:\phi_{\nu_{\tau}}^0=0:1:0.
  \end{equation}
  On the other hand, the decays of neutrons from the heavy nuclei photo-dissociation provide a pure $\bar{\nu}_e$ generation to bring the ratios in the neutron-beam case~\cite{pure electron}:
  \begin{equation}
  \phi_{\nu_e}^0:\phi_{\nu_{\mu}}^0:\phi_{\nu_{\tau}}^0=1:0:0.
  \end{equation}
 The flavor ratio detection can provide us information concerning the UHECN sources.

 Since the neutrinos are mixing among the mass eigenstates in the flavor eigenstates, the neutrino flavor ratios change during propagating and the final ratios at Earth are different from the initial ratios at sources. In the three-generation neutrino framework, the neutrino mixing is well described by a $3\times3$ unitary matrix $U$ which is known as the Pontecorvo-Maki-Nakawaga-Sakata~(PMNS) mixing matrix~\cite{PMNS1,PMNS2}. The flavor mixing of three neutrino generations is written as
\begin{equation}
|\nu_{l}\rangle=\sum_jU_{lj} |\nu_j\rangle,
\label{mixing}
\end{equation}
where the subscripts $l=e,\mu,\tau$ denote the flavour eigenstates and $j=1,2,3$ represent the mass eigenstates. In the standard parametrization,
the PMNS matrix is expressed by three mixing angles $\theta_{12}, \theta_{13}, \theta_{23}$ and one CP-phase angle $\delta$ in a form
\begin{widetext}
\begin{equation}
V(\theta_{12},\theta_{23},\theta_{13},\delta)=
\left(\begin{array}{ccc}
            c_{12}c_{13} & s_{12}c_{13} & s_{13}\e^{-\i\delta}\\
           -c_{12}s_{23}s_{13}\e^{\i\delta}-s_{12}c_{23}& -s_{12}s_{23}s_{13}\e^{\i\delta}+c_{12}c_{23}& s_{23}c_{13}\\
             -c_{12}c_{23}s_{13}\e^{\i\delta}+s_{12}s_{23}& -s_{12}c_{23}s_{13}\e^{\i\delta}-c_{12}s_{23}& c_{23}c_{13}\\

      \end{array}\right),
      \label{PMNS}
      \end{equation}

\end{widetext}
where $s_{ij}$ and $c_{ij}$ denote $\sin \theta_{ij}$ and $\cos \theta_{ij}~(i, j=1,2,3)$.  Here we use the recently updated global fit data~\cite{global fit} listed in Table~\ref{tab:1}.
\begin{table}[]
      \caption{The global fit of neutrino mixing parameters~\cite{global fit}, with NH or IH denoting the normal or inverted hierarchy of mass eigenstates. }
      \label{tab:1}
      \centering
      \begin{tabular}{cll}
      \hline
      \hline
        \noalign{\vspace{0.5ex}}
       parameter & best fit$\pm 1\sigma$\  & $3\sigma$ range\\
             \noalign{\vspace{0.5ex}}
      \hline
      \noalign{\vspace{0.5ex}}
      $\sin^2\theta_{12}({\rm NH\ or\ IH})$ & $0.304_{-0.012}^{+0.013}$ & $0.270 \to 0.344$ \\
      \noalign{\vspace{0.5ex}}
      $\sin^2\theta_{23}({\rm NH})$ & $0.452_{-0.028}^{+0.052}$ & $0.382 \to 0.643$ \\
      \noalign{\vspace{0.5ex}}
      $\sin^2\theta_{23}({\rm IH})$ & $0.579_{-0.025}^{+0.037}$ & $0.389 \to 0.644$ \\
      \noalign{\vspace{0.5ex}}
      $\sin^2\theta_{13}({\rm NH})$ & $0.0218_{-0.0010}^{+0.0010}$ & $0.0186 \to 0.0250$ \\
      \noalign{\vspace{0.5ex}}
      $\sin^2\theta_{13}({\rm IH})$ & $0.0219_{-0.0011}^{+0.0010}$ & $0.0188 \to 0.0251$ \\
      \noalign{\vspace{0.5ex}}
      $\delta/\deg({\rm NH})$ & $306^{+39}_{-70}$ & $0 \to 360$\\
       \noalign{\vspace{0.5ex}}
      $\delta/\deg({\rm IH})$ & $254^{+63}_{-62}$ & $0 \to 360$\\
        \noalign{\vspace{0.5ex}}
      \hline
      \hline

      \end{tabular}
\end{table}

\begin{figure}[h]


  \centering
 \includegraphics[width=\linewidth]{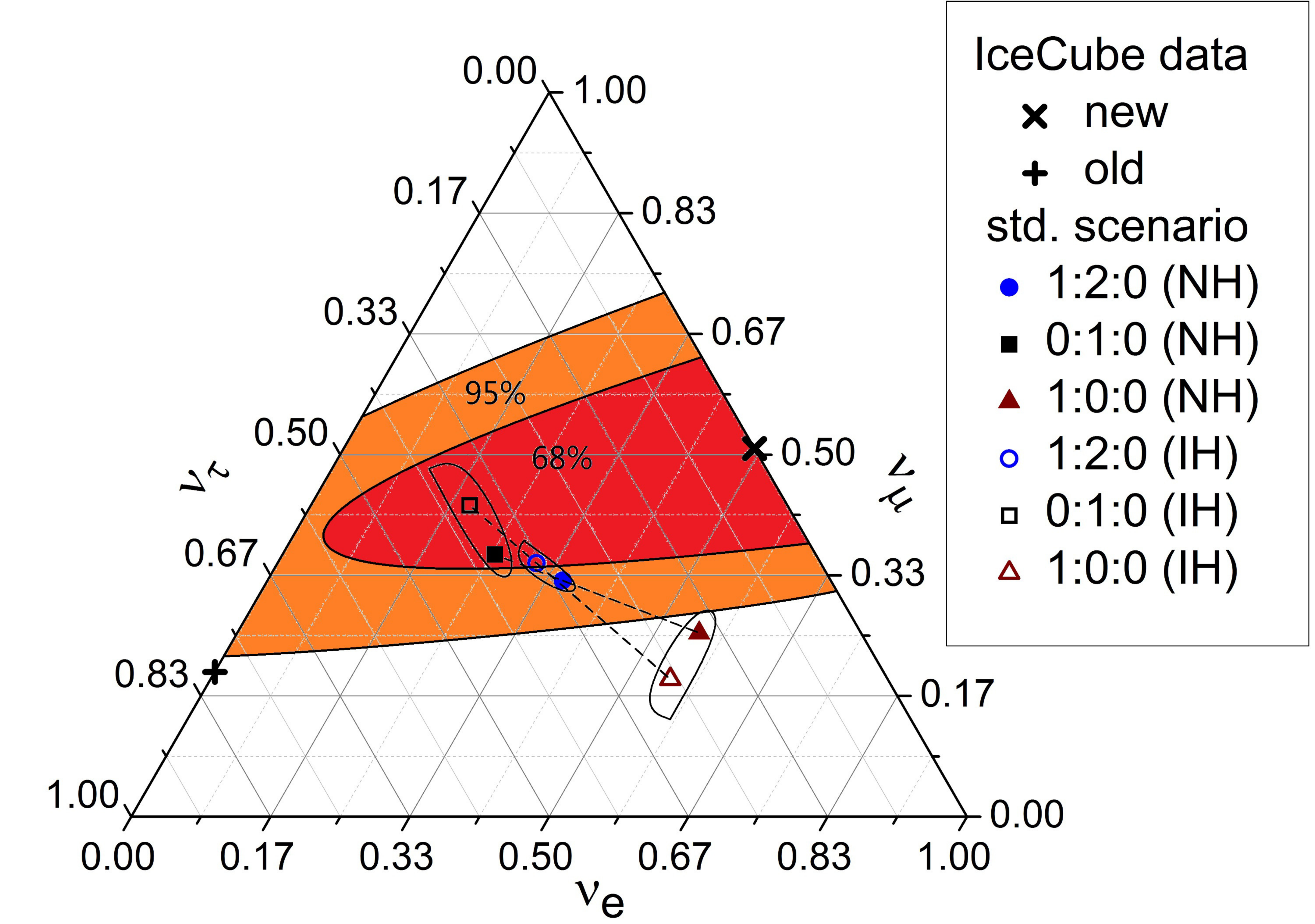}

\caption{The comparison of the predicted flavor ratios at Earth in the standard oscillation scenario with the IceCube data. The NH cases are marked by the solid points corresponding to the initial ratios by shapes, and the IH cases are marked by the open points.  The regions marked by the solid lines represent the error ranges due to the $3\sigma$ range of mixing parameters. The 68\% and 95\% confidence regions of the IceCube data are also indicated. }
\label{fig:1}
    \end{figure}

For an individual neutrino arisen at a determined position, the transition probability is related to the mass square difference and the propagation length.  But if performing an average over all the neutrinos arisen at arbitrary positions, only mixing parameters are related to the ratio changes. The flavor ratio detected at Earth $\phi_\alpha^\oplus$ is written as~\cite{pion decay 2}
\begin{equation}
\phi_l^\oplus=\sum_{l^\prime, i} |U_{l i}|^2|U_{l^\prime i}|^2\phi_{l^\prime}^0.
\label{standard formula}
\end{equation}
If considering a tribimaximal mixing case, the flavor ratios in Eq.~(\ref{initial ratio}) originated from the pion-decay sources change into~\cite{pion decay 2}
 \begin{equation}
 \phi_{\nu_e}^\oplus:\phi_{\nu_{\mu}}^\oplus:\phi_{\nu_{\tau}}^\oplus=\frac{1}{3}:\frac{1}{3}:\frac{1}{3}.
 \end{equation}
Using the global fitting values of  mixing parameters, we can evaluate the final detected flavor ratios of different initial ratios discussed above. The results are shown in Table~\ref{tab:2} and Fig.~\ref{fig:1}. The ratios in a determinate mass hierarchy are in a line, because the ratios at source are linear correlated, i.e.,
 \begin{equation}
 (1:2:0)=(1:0:0)+2\times(0:1:0).
 \end{equation}
 The ranges marked by the solid lines in Fig.~\ref{fig:1} represent the error ranges due to the $3\sigma$ range of the mixing parameters. Comparing with the latest data from the IceCube experiment~\cite{IceCube new1, IceCube new2}, we find that in both NH and IH cases, the pion-decay sources and muon-damped sources are compatible with the IceCube data, but the neutron-beam sources are disfavored even thought we broaden the error range to $3\sigma$. Since the flavor ratios in different mass hierarchy cases have distinct deviations, it is expected that the future data can put strong constraint on the mass hierarchy in the standard oscillation scenario.

\begin{table}[]
      \caption{The predicted flavor ratios of neutrinos at the Earth in different cases and initial ratios. The ratios are in the $\nu_e:\nu_{\mu}:\nu_{\tau}$ order. The best-fit data of the IceCube experiment can be found in Refs.~\cite{IceCube new1, IceCube new2}}
      \label{tab:2}
      \centering
      \begin{tabular}{ccl}
      \hline
      \hline
        \noalign{\vspace{0.5ex}}
       cases & ratios at source & ratios at the Earth\\
      \hline
      \noalign{\vspace{0.5ex}}
        & 1 : 2 : 0 & 0.35 : 0.33 : 0.32 (NH) \\

        &   & 0.31 : 0.35 : 0.34  (IH) \\
      \noalign{\vspace{0.5ex}}
    standard  & 0 : 1 : 0 & 0.25 : 0.36 : 0.39 (NH) \\

     oscillation&   & 0.19 : 0.43 : 0.38 (IH) \\
      \noalign{\vspace{0.5ex}}
        & 1 : 0 : 0 & 0.55 : 0.25 : 0.20 (NH) \\

      &   & 0.55 : 0.19	: 0.26 (IH) \\
            \noalign{\vspace{0.5ex}}
         \hline
      \noalign{\vspace{0.5ex}}
      decay (NH)& &0.68 :	0.21 :	0.11\\
        \noalign{\vspace{0.5ex}}
        decay (IH)& &0.02 : 0.57 : 0.41\\
              \noalign{\vspace{0.5ex}}
        \hline
        \noalign{\vspace{0.5ex}}
         &1 : 2 : 0&0.50 :	0.27 : 0.23\\
        \noalign{\vspace{0.5ex}}
       deacy&0 : 1 : 0&0.47 : 0.28 :	0.25\\
        \noalign{\vspace{0.5ex}}
        (degenerate)&1 : 0 : 0&0.56 : 0.25	: 0.19\\
         \noalign{\vspace{0.5ex}}
        \hline
        \noalign{\vspace{0.5ex}}
        IceCube &&0.49 : 0.51 : 0 (new)\\
        \noalign{\vspace{0.5ex}}
       best-fit&&0 : 0.2 : 0.8 (old)\\
         \noalign{\vspace{0.5ex}}
       \hline
      \hline
      \end{tabular}
\end{table}

    \begin{figure}[h]


  \centering
 \includegraphics[width=\linewidth]{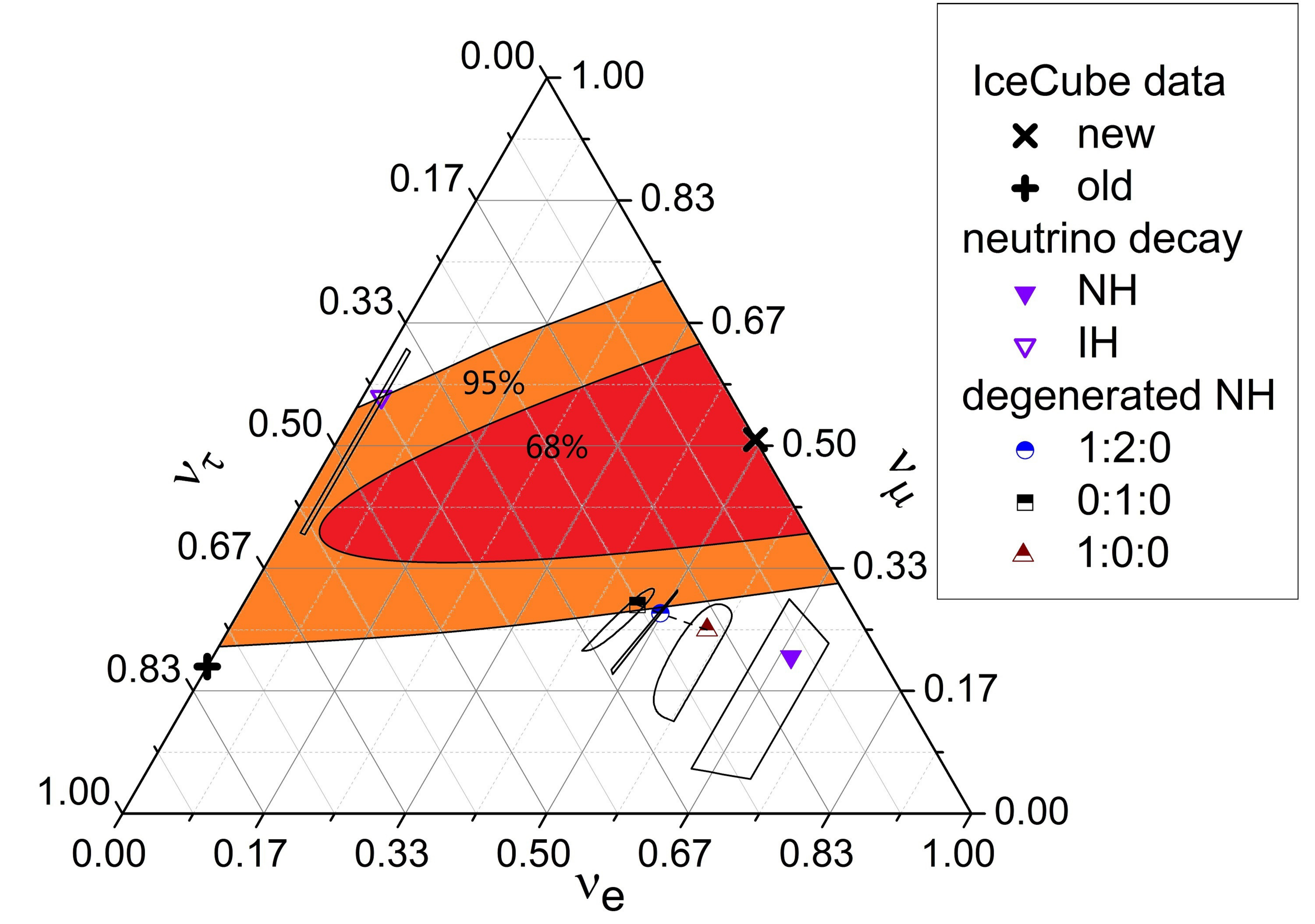}

\caption{The comparison of the predicted flavor ratios in the the neutrino decay scenario with the IceCube data. The non-degenerated cases in NH and IH are marked by the inverted triangles. The points of the degenerated cases from different sources are in a line, for the similar reason of the standard oscillation scenario in Fig.~\ref{fig:1}.}
\label{fig:2}
    \end{figure}

Now we consider a new scenario beyond the standard model. Since the splitting among three mass eigenstates have been detected, the heavier neutrinos may decay into the lighter ones via flavor changing processes~\cite{neutrino decay 1, neutrino decay 3, neutrino decay 4}. The dominate neutrino decay manner is thought to be
\begin{equation}
\nu_i\rightarrow\nu_j+X \   {\textrm{ and} } \    \nu_i\rightarrow\bar{\nu}_j+X,\label{decay}
\end{equation}
where $\nu_i ~ (i=1,2,3)$ are neutrino mass eigenstates and $X$ is a very light or massless particle, e.g., a Majoron~\cite{astrophysical neutrino decay}. Many more detailed discussions have been proposed~\cite{astrophysical neutrino, Maltoni:2008jr, Bi:2009de, Wang:2013bvz, Lai:2013isa}.     Since the exiting limit on the half-life $\tau$ of the decay channel in Eq.~(\ref{decay}) is too weak to eliminate the UHECN decay with high energy and lengthy propagation, i.e., $L\gg\tau$, only the lightest neutrino can arrive at Earth.  Thus, the final flavor ratios detected at Earth depend on the mass hierarchy and the mixing parameters. The mass hierarchy determines the remain mass eigenstate, and the mixing parameters determine the flavor components of each mass eigenstate.
If the neutrino mass eigenstates are in normal hierarchy (NH), i.e., $m_3>m_2>m_1$, the UHECN beam contains the only stable mass state $\nu_1$. So that Eq.~(\ref{standard formula}) reduces to
 \begin{equation}
\phi_l^\oplus=|U_{l 1}|^2.
\end{equation}
 The flavor ratios at Earth are
  \begin{equation}
  \phi_{\nu_e}^\oplus:\phi_{\nu_{\mu}}^\oplus:\phi_{\nu_{\tau}}^\oplus
  =0.68^{+0.03}_{-0.04}:0.21^{+0.08}_{-0.16}:0.11^{+0.17}_{-0.06}.\label{NH ratio}
  \end{equation}
The $3\sigma$ range of the mixing parameters is considered in the error calculation. A probable ratio is an arbitrary but unitary combination of the three $\phi_l$'s in the error range, as shown in Fig.~\ref{fig:1}.
If the neutrino mass eigenstates are inverted hierarchy (IH), i.e., $m_2>m_1>m_3$, the remain mass eigenstate turns to be $\nu_3$. Hence the flavor ratios at Earth change into
 \begin{equation}
  \phi_{\nu_e}^\oplus:\phi_{\nu_{\mu}}^\oplus:\phi_{\nu_{\tau}}^\oplus
  =0.02^{+0.003}_{-0.003}:0.57^{+0.06}_{-0.19}:0.41^{+0.019}_{-0.06}.\label{IH ratio}
  \end{equation}

Further more, we consider the neutrino decay in degenerate cases. Since the mass splitting between $\nu_1$ and $\nu_2$, i.e., $\Delta m_{21}^2$, is two orders of magnitude smaller than $\Delta m_{31}^2$ and $\Delta m_{32}^2$~\cite{MINOS,Daya Bay,T2K new}, it is possible to conjecture that the mass eigenstates $\nu_1$ and $\nu_2$ are approximatively degenerate relative to $\nu_3$~\cite{double beta decay}. This assumption is supported by the experiments of the neutrino-less double beta decay~\cite{double beta decay} and the cosmological observations~\cite{CB}, since the mass splitting $\Delta m_{21}^2$ is much smaller than the bounds of absolute mass scales.  For the inverted hierarchy, both $\nu_1$ and $\nu_2$ decay to $\nu_3$, so still only the $\nu_3$ is stable. Hence, the degenerate IH case leads to the same result as that of the non-degenerate IH case. But in the normal hierarchy case, $\nu_3$ is the heaviest mass eigenstate, so both $\nu_1$ and $\nu_2$ are stable due to the degeneration. Each stable mass eigenstate originates from not only the initial flavor eigenstates but also the $\nu_3$ decay. Since the flavor components of $\nu_1$ and $\nu_2$ are disparate, i.e., $U_{l1}\neq U_{l2}$, and the branching ratios of the $\nu_3$ decay may be different, the latter two factors in Eq.~(\ref{standard formula}) can not be cancelled. That is to say, the final flavor ratios at the Earth are related to the initial flavor ratio source and the branching ratios. As a straightforward assumption, i.e., letting the branching ratios of $\nu_1$ and $\nu_2$ both being equal to 50\%, Eq.~(\ref{standard formula}) changes to
\begin{equation}
\phi_l^\oplus=\sum_{l^\prime, i}|U_{li}|^2(|U_{l^\prime i}|^2+|U_{l^\prime3}|^2/2)\phi_{l^\prime}^0,\label{degenerate fomular}
\end{equation}
where $i=1, 2$ and $l'=e, \mu, \tau$. Then we can get different final flavor ratios for different sources.

The results of neutrino decay scenarios are shown in Table~\ref{tab:2} and Fig.~\ref{fig:2}. The $3\sigma$ error regions of degenerate NH cases are smaller than those of non-degenerate cases, because the flavor ratios in Eq.~(\ref{degenerate fomular}) are more complicated and the errors from different mixing parameters cancel each other, due to the unitary of the PMNS matrix. The error region of the IH case is narrower along the $\nu_e$ direction, because the error range of $\theta_{13}$ is much smaller than those of other mixing parameters. We can find that the IH case (only $\nu_3$ is stable) and the degenerate NH case from the pion-decay sources (1:2:0) and the muon-damped sources (0:1:0) are compatible with the IceCube data at the $3\sigma$ error range, although the best-fit points in these cases are at the boundary of 95\% confidence region.  The non-degenerate NH case and the degenerate NH case from the neutron-beam sources (1:0:0) are disfavored comparing with the latest IceCube data.

In the early years, the flavor ratios of UHECNs were discussed on a premise that the matrix element $U_{13}$ is vanishing~\cite{astrophysical neutrino decay, astrophysical neutrino}. After the discovery of the  non-vanishing mixing angle $\theta_{13}$~\cite{e9}, many theoretical calculations based on the updated mixing parameters are proposed~\cite{Lai:2010fq, Fu:2012zr, Xu:2014via, Chen:2014gxa, Palladino:2015zua, Palladino:2015vna, Arguelles:2015dca, Bustamante:2015waa, Pagliaroli:2015rca}, in both standard and non-standard scenarios. According to the new result announced by the IceCube collaboration~\cite{IceCube new2}, two kinds of sources in the standard oscillation scenario, i.e.,  pion-decay sources and muon-damped sources, are well compatible with the new IceCube data, while the non-standard oscillation scenarios have not been found. We propose a new probable case that the mass eigenstates $m_1$ and $m_2$ are degenerate in the neutrino decay scenario. The degeneracy of mass eigenstates changes the stable states, and results in different final flavor ratios from different sources. In the degenerate NH case, we find that the theoretical predictions of the neutrino decay scenario are still permissible comparing with the IceCube data, by using the updated global fitting mixing parameters.

\begin{figure}[h]


 \includegraphics[width=\linewidth]{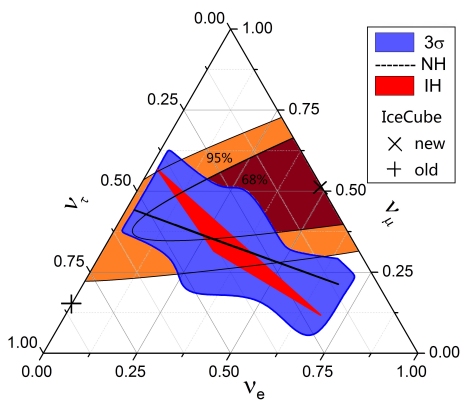}

\caption{The comparison of the predicted flavor ratios in the arbitrary initial ratio scenario with the IceCube data. The best fit regions in NH and IH are the line shape region and the thin triangle in the middle area. The $3\sigma$ region is denoted by the filled area. The IceCube best fit points, 68\% and 95\% confidence regions are also shown as comparisons. }
\label{fig:3}
    \end{figure}

In the discussions above, the initial flavor ratio at the source was introduced as a constraint. Different initial ratios lead to different final ratios that are detectable by the IceCube experiment. Hence, one can conjecture which source or sources play the major roles according to the IceCube data or which way of the neutrino decay is probable, according to the IceCube data. On the other hand, if all the possible initial ratios are included, i.e., the initial ratios are arbitrary, one can also get a constraint on the final ratios by the restrict on the neutrino mixing and oscillation only. Here we consider a general example that not only flavor ratio is arbitrary, even the ratio of mass eigenstates is also arbitrary, which means the mass eigenstates are allowed to be incoherent during propagating. The later case is more general since it can completely cover the region of the former case. Fig.~\ref{fig:3} shows the final ratios with the mixing parameters best fit regions and the $3\sigma$ regions in both NH and IH case, compared with the IceCube data. The $3\sigma$ regions in NH and IH case are almost the same. In Fig.~\ref{fig:3}, the best fit regions makes up only a very small portion in the ratio triangle, no matter in NH or IH case. It suggests that the neutrino mixing framework can actually provide a strong constraint on the flavor ratio detected at Earth, no matter what happened at the source.

Another point can not be neglected is that the best fit points of the IceCube data are not covered by the general $3\sigma$ region, though the 95\% and 68\% confidence regions overlap with the $3\sigma$ region in some areas. That may indicate the probability of new physics models beyond the the neutrino mixing and oscillation framework today. It is well expected that the UHECNs detections can be used as constraints of the mixing parameters and the mechanism.

In summary, we discuss the flavor ratios of UHECNs originated from three possibly astrophysical sources. The final flavor ratios at the Earth depend on the initial ratios because of the neutrino oscillation or the neutrino decay. By using the recently updated global fitting results of mixing parameters, we evaluate the final flavor ratios within the $3\sigma$ error range, and compare our results with the recent IceCube data~\cite{IceCube new1, IceCube new2}. In the standard oscillation scenario, both NH and IH cases from the pion-decay sources (1:2:0) and the muon-damped sources (0:1:0) are compatible with the IceCube data, but the neutron-beam sources (1:0:0) are disfavored for both NH and IH cases. If considering the neutrino decay, we find that the IH case and the degenerate NH case from the 1:2:0 and 0:1:0 sources are permissible within the $3\sigma$ error range, while the non-degenerate NH case or the degenerate NH case from the 1:0:0 sources are disfavored with the data. If removing all the other restricts but neutrino mixing and oscillation, we can also get a constraint on the final ratios. Since the difference between NH and IH cases are considerably significant in the standard scenario, the decay scenario and the general constraint, the future data of UHECNs can provide us more information on the neutrino properties.

We acknowledge the stimulating conversations with Pisin Chen. This work is supported by the National Natural Science Foundation of China (Grants No.~11120101004 and No.~11475006).


\end{document}